\documentclass[twocolumn,showpacs,preprintnumbers,amsmath,amssymb]{revtex4}

\usepackage{amsfonts} 

\usepackage{hyperref}

\usepackage{color} 
 
\usepackage[utf8,applemac]{inputenc} 
\usepackage{tensor}
\usepackage{ulem}
\usepackage{tikz}
\usepackage{graphicx}
\usepackage{subfig}	

\usetikzlibrary{positioning}
\usetikzlibrary{calc}

\newcommand{\bea}{\begin{eqnarray}}
\newcommand{\eea}{\end{eqnarray}}
\newcommand{\be}{\begin{equation}}
\newcommand{\ee}{\end{equation}}
\def\nn{\nonumber}

\def\p{\partial}
\def\eps{\epsilon}

\usepackage{mathtools}

\begin{document}
\title{Transition from inspiral to plunge into a highly spinning black hole}

  \author{Geoffrey Comp\`ere$^\dagger$, Kwinten Fransen$^*$ and Caroline Jonas$^\dagger$}
\affiliation{%
\textit{$^\dagger$ Centre for Gravitational Waves, Universit\'{e} Libre de Bruxelles, 
 International Solvay Institutes, CP 231, B-1050 Brussels, Belgium \\
 $^*$Centre for Gravitational Waves, Institute for Theoretical Physics, \\
KU Leuven, Celestijnenlaan 200D, 3001 Leuven, Belgium}
}

\begin{abstract}
We extend the Ori-Thorne-Kesden procedure to consistently describe the non-quasi-circular transition around the ISCO from inspiral to plunge into a black hole of arbitrary spin, including near-extremal. We identify that for moderate or high spins the transition is governed by the Painlev\'e transcendent equation of the first kind while for extremely high spins it is governed by a self-similar solution to the Korteweg-de Vries equation. We match the transition solution at leading order in the high spin limit with the analytical quasi-circular inspiral in the near-horizon region. We also show that the central black hole of an extreme mass ratio binary has a near-extremality parameter that scales at least as the mass ratio due to superradiant gravitational wave emission, which excludes extremely high spins.

\end{abstract}


\maketitle

A theoretical modeling effort is required in order to produce a database of accurate and faithful templates for extreme mass ratio inspirals (EMRIs) for the LISA mission \cite{Audley:2017drz,AmaroSeoane:2007aw} or its proposed extension \cite{Baibhav:2019rsa}. In addition to the main science objectives of the mission, intermediate mass ratio coalescences (IMRACs) which consist in intermediate mass black holes plunging into supermassive black holes  constitute a potential source for LISA \cite{Miller:2004va}. Such sources require an accurate modeling of the transition from inspiral to plunge since the number of cycles spent in that latter phase is observationally significant \cite{Smith:2013mfa}. 

The transition from inspiral to plunge was modeled for quasi-circular inspirals in the equatorial plane by Ori and Thorne \cite{Ori:2000zn} within black hole perturbation theory under some simplifying assumptions (see also the EOB framework \cite{Buonanno:2000ef} and \cite{Buonanno:2005xu,Damour:2007xr,Sundararajan:2008bw,Damour:2009kr,Kesden:2011ma,Pan:2013rra,Taracchini:2014zpa,Apte:2019txp} for extensions). Such a transition occurs around  the innermost stable circular orbit (ISCO). One of the original assumptions of  \cite{Ori:2000zn} was that  the orbit is quasi-circular around the transition. However, such an assumption was shown to lead to mathematical inconsistencies by Kesden \cite{Kesden:2011ma}, though its analysis has been overlooked in the subsequent literature. It is therefore required to relax that hypothesis and consider a non-circular transition motion. The Ori-Thorne-Kesden model applies for all moderate spins but becomes inconsistent in the high spin regime where $\lambda \equiv \sqrt{1-J^2/M^4} \rightarrow 0$ \cite{Kesden:2011ma}. The main purpose of this paper is to complete the Ori-Thorne-Kesden analysis to cover the high spin regime where new qualitative features arise. 

Geometrically thin disks allow to spin up black holes only up to the Thorne bound $\lambda \geq 0.06$ \cite{1974ApJ...191..507T}. Other accretion models might however by-pass this bound since no fundamental limitation exists on how fast accretion can spin up a black hole \cite{1980AcA....30...35A}. More fundamentally, the high spin limit $\lambda \rightarrow 0$ can be viewed as the leading order result of a perturbative expansion for small $\lambda$. In the high spin regime, the ISCO lies within the near-horizon region of Kerr which is described by the near-horizon geometry of extremal Kerr (NHEK) \cite{Bardeen:1999px} up to $O(\lambda^{1/3})$ corrections and corrections due to the self-force of the incoming compact object. At leading order in the high spin limit $\lambda \rightarrow 0$ and neglecting the self-force the physics around the ISCO is exactly described by physics in the NHEK geometry with appropriate boundary conditions that relate the near-horizon region to the exterior asymptotically flat region \cite{Porfyriadis:2014fja,Hadar:2014dpa,Hadar:2015xpa,Gralla:2015rpa,Gralla:2016qfw,Hadar:2016vmk,Compere:2017hsi,Chen:2019hac}. 

\begin{figure}\label{fig:matching}
	\includegraphics[width=.50\textwidth]{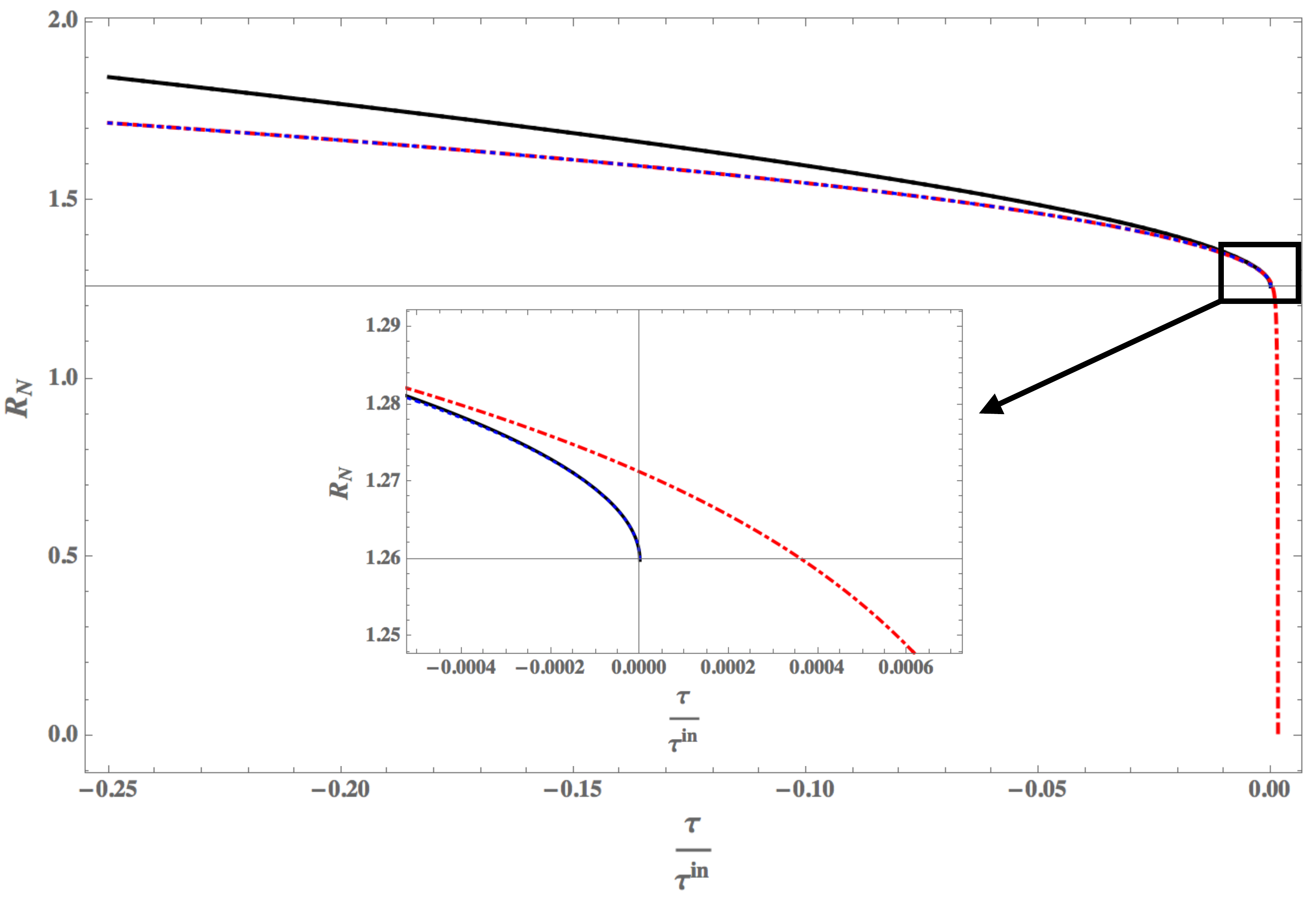}\vspace{-10pt}
	\caption{Evolution of the near-horizon radius $R_N$ in terms of proper time $\tau$ during the inspiral (solid black line) and transition to plunge (dotted-dashed red line) for a binary with $\lambda = 10^{-3}$ and mass ratio $\eta = 10^{-6}$. The ISCO lies at $R_N=2^{1/3}$. The final plunge occurs shortly after reaching the critical angular momentum at $\tau = \tau_* = 0$ and is therefore described by a subcritical ($\ell < \ell_*$) geodesic.}\vspace{-10pt}
\end{figure}

The boundary conditions at the entry of  the near-horizon geometry can be deduced from the late inspiral in the exterior near-extremal Kerr geometry. In the equatorial plane, an eccentric inspiral tends to circularize \cite{1996PhRvD..53.3064R} up to a critical radius where the eccentricity start increasing \cite{Kennefick:1998ab}. Now, that critical radius lies within the NHEK region in the high spin limit though it remains distinct from the ISCO \cite{Kennefick:1998ab}. If the rate of circularization is sufficiently fast, we can therefore assume a quasi-circular inspiral entering the near-horizon region, where the non-quasi-circular transition takes place. Such a quasi-circular inspiral was analytically obtained in \cite{Gralla:2016qfw}. For LISA sources, the rate of circularization is expected to be such that eccentricity and inclination will generally be present at the separatrix between bound and unbound motion \cite{Glampedakis:2002ya}. We will only address in the following the transition from a quasi-circular inspiral and leave the study of the general transition for further work.

At the end of the transition, the radiation reaction becomes negligible for an EMRI and the motion is geodesic. The final transition motion is therefore  described at leading order in  small mass ratio $\eta \to 0$ and in the high spin limit $\lambda \rightarrow 0$ by a geodesic plunge in NHEK. The classification of such trajectories and a derivation of the associated Teukolsky waveforms was obtained in \cite{Compere:2017hsi} (see also \cite{Kapec:2019hro,Compere:2019bb}). In this context, it is natural to ask what the final parameters of the geodesic plunge are as a function of $\lambda \ll 1$ and of the mass ratio $\eta \ll 1$. 

The main result of this paper is the description of the motion of a point particle probe from the quasi-circular inspiral through the transition and up to the final plunge, at leading order in the high spin limit of the central black hole. This motion is summarized on Figure 1. We will distinguish the standard high spin case $\eta \ll \lambda$, the marginal high spin case $\eta \sim \lambda$ and the extremely high spin case $\lambda \ll \eta$.  We will show that the standard high spin case leads to a unique matching solution.  On the contrary, the marginal and extremely high spin cases will not match the quasi-circular inspiral. Moreover,  the extremely high spin case will be shown to be inconsistent with the spin evolution of the central black hole. 

\vspace{10pt}
\noindent \textit{Note Added.} In the final stages of this work, we became aware of overlapping results that were independently obtained in \cite{Simon:2019aa}.

\section{General features of the transition}
\label{sec:hyp}

In the following we consider the probe angular momentum $\ell = \frac{L}{\mu M}$ per unit probe mass $\mu$ and rescaled by the central black hole mass $M$, and the probe energy $\tilde e= \frac{E}{\mu}$ per unit probe mass with respect to the Boyer-Lindquist asymptotic timelike Killing vector $\p_{\tilde t}$. We denote as $\tilde r= r/M$ the adimensional Boyer-Lindquist radial coordinate of Kerr, $\tilde a = a/M$ and the near-extremality parameter $\lambda = \sqrt{1-\tilde a^2}$.

Following Ori-Thorne \cite{Ori:2000zn} and subsequent work, we assume the following three simplifying hypotheses: 
\begin{itemize}
\item The decrease rate of probe angular momentum $\ell$ per unit dimensionless proper time $\tau$ (proper time divided by $M$) is equal to the rate of a probe on a circular orbit at the ISCO. 
\item We neglect the radial self-force.
\item The proper time ticks as along a circular orbit.
\end{itemize}
The first hypothesis is equivalent to a linear decay rate of the probe angular momentum, 
\bea
\ell(\tau) =\ell_* -\kappa_* \eta (\tau - \tau_*),\qquad \kappa_* \equiv \frac{8 \sigma_* }{\sqrt{3}}.  \label{linl}
\eea
Here $\ell_*$ is the probe angular momentum per unit $M \mu$ at the ISCO, $\eta$ is the mass ratio, $\sigma_*= \sigma_{*,\infty} + \sigma_{*,H}$ is a constant determined by the total angular momentum flux emitted from a circular orbit at the ISCO reaching infinity and the horizon, and $\tau_*$ is the proper time at which one would reach $\ell = \ell_*$. For the Schwarzschild black hole, $\sigma_* \approx 0.004$  \cite{Ori:2000zn} while in the extremely high spin limit \cite{Gralla:2016qfw}, 
\bea
\sigma_{*,\infty} \approx 0.987, \quad \sigma_{*,H} \approx -0.133.\label{sigma}
\eea
The limitations of the second hypothesis were discussed in \cite{Pound:2005fs}. 

 During the transition, we do \emph{not} assume quasi-circularity, as it was shown to be inconsistent \cite{Kesden:2011ma}. In \cite{Ori:2000zn} the decrease rate of probe energy per proper time was fixed in terms of the corresponding decrease of probe angular momentum after assuming quasi-circularity. Here, the decrease rate of probe energy is fixed by requirement that a potential exists for the radial motion. More precisely, assuming no radial self-force, the first and second order radial geodesic equations take the form
\bea
\left( \frac{d\tilde r}{d\tau} \right)^2 &=& \tilde e^2 - \tilde V(\tilde r , \tilde e, \ell) ,\label{geoR} \\
\frac{d^2\tilde  r}{d\tau^2} &=& -\frac{1}{2}\frac{\p \tilde V(\tilde r , \tilde e, \ell)}{\p \tilde r},\label{accR}
\eea
where the radial potential is given by
\bea
\tilde V(\tilde r,  \tilde e, \ell)\! &=&\! 1- \frac{2}{\tilde r} - \frac{2 (\ell- \tilde a  \tilde e)^2}{\tilde r^3} - \frac{\tilde a^2 ( \tilde e^2 - 1) - \ell^2 }{\tilde r^2}.
\eea
These equations are compatible given the following constraint equation is obeyed
\bea
\frac{d \tilde e}{d\tau} \frac{\p (\tilde V-\tilde e^2 )}{\p  \tilde  e} + \frac{d \ell}{d\tau} \frac{\p \tilde V}{\p \ell} = 0.  
\eea
We impose it as the evolution of the probe energy after substituting \eqref{linl}, 
\bea
\frac{d \tilde e}{d\tau} = \kappa_* \eta \frac{\p \tilde V/\p \ell}{\p (\tilde V - \tilde e^2)/ \p  \tilde e}. \label{geo2R}
\eea

In order to describe the transition around the ISCO we define the deviation variables
\bea
\tilde R &\equiv& \tilde r - \tilde r_{*}, \\
\chi &\equiv& \tilde \Omega_*^{-1} (\tilde e -\tilde  e_* ),\\
\xi &\equiv& \ell - \ell_*,
\eea
where all quantities $\tilde r_{*}$, $\tilde \Omega_*$, $\tilde e_*$, $\ell_*$ at the ISCO are detailed in Appendix \ref{app:circ}. At the ISCO, we have
\bea
\frac{\p^2 \tilde V}{\p \tilde r^2}|_*  = \tilde  e_*- \frac{1}{2} \frac{\p \tilde V}{\p  \tilde e}|_* - \frac{1}{2 }\tilde \Omega_*^{-1} \frac{\p \tilde V}{\p \ell}|_*= 0.
\eea
We now Taylor expand the potential $\tilde V$ at order $O(\tilde R^4)$ and we truncate to linear order in the deviation parameters $\xi$, $\chi$. The radial equation \eqref{geoR} and energy evolution equation \eqref{geo2R} become 
\bea
\frac{d^2 \tilde R}{d\tau^2} &=& -\alpha_* \tilde R^2 + (\gamma_* \tilde R +\beta_*)\xi-\frac{1}{2}(\chi-\xi)[c_* \tilde R+d_*],\nonumber \\
\frac{d(\chi - \xi)}{d\tau} &=&\kappa_* \eta \frac{\gamma_* \tilde R^2 + 2\beta_* \tilde R}{-\frac{1}{2} c_*\tilde R^2-d_*\tilde R+\delta_*},\label{master}
\eea
where the Taylor coefficients are defined as 
\bea
\alpha_* &=& \frac{1}{4} \frac{\p^3 \tilde V}{\p \tilde r^3}|_*,\quad c_* = \tilde \Omega \frac{\p^3 \tilde V}{\p \tilde r^2 \p \tilde  e}|_*, \\
\beta_* &=&\!\! - \frac{1}{2} \left( \frac{\p^2 \tilde V}{\p \tilde r \p \ell} +\tilde \Omega \frac{\p^2 \tilde V}{\p \tilde r \p\tilde   e} \right)|_*,\; d_* = \tilde \Omega \frac{\p^2 \tilde V}{\p \tilde r \p  \tilde e}|_*,\\
\gamma_* &=& -\frac{1}{2} \left( \frac{\p^3 \tilde V}{\p \tilde r^2 \p \ell} +\tilde \Omega \frac{\p^3 \tilde V}{\p \tilde r^2 \p \tilde e} \right)|_*,\quad \delta_*=\frac{\p \tilde V}{\p \ell}|_*.
\eea
These equations govern the transition around the ISCO for arbitrary spins. We will check below that all terms neglected in this Taylor expansion asymptote to zero in an appropriate region around the ISCO.

\section{The Ori-Thorne-Kesden equations}

The observation of Ori-Thorne \cite{Ori:2000zn} and Kesden \cite{Kesden:2011ma} is that for standard spins the transition equations \eqref{master} are consistent with the scaling 
\bea
\tau -\tau_* & \sim &\eta^{-1/5},\qquad \tilde R \sim \eta^{2/5},\nonumber \\
\chi &\sim  &\xi \sim \eta^{4/5},\qquad \chi - \xi \sim \eta^{6/5}. \label{scaleOTK}
\eea
Assuming that scaling, the Taylor terms multiplying $\gamma_*$, $c_*$ and $d_*$ are subleading in the small mass ratio limit $\eta \rightarrow 0$  and can be neglected. At leading order in $\eta$ the equations reduce to 
\bea
\frac{d^2 \tilde R}{d\tau^2} &=& - \alpha_* \tilde R^2  -\kappa_* \eta \beta_* (\tau - \tau_*), \\
\frac{d(\chi - \xi)}{d\tau} &=& 2 \kappa_* \eta \frac{\beta_*}{\delta_*} \tilde R,
\eea
after using \eqref{linl}. One can set these equations in normalized form including the order of the subleading correction
\bea
\frac{d^2X}{dt^2} =-X^2 - t + O(\eta^{2/5}),\quad \frac{dY}{dt}=2X+ O(\eta^{2/5}),\label{OTK}
\eea
after defining
\bea
\tilde R &=& \frac{(\beta_* \kappa_* \eta)^{2/5}}{\alpha_*^{3/5}}X(t),\\
\tau&=& \tau_*+(\alpha_* \beta_* \kappa_* \eta)^{-1/5} t, \\
\chi - \xi &=& \frac{(\beta_* \kappa_* \eta)^{6/5}}{\alpha_*^{4/5} \delta_*} Y(t).
\eea
By construction, these equations are consistent with the first order radial equation
\bea
(\frac{dX}{dt})^2 = -\frac{2}{3}X^2 - 2 X t +Y+ O(\eta^{2/5}).
\eea
The first equation of \eqref{OTK} is the Ori-Thorne equation. We identify it here as the Painlev\'e transcendent equation of the first kind. It is a typical equation in bifurcation phenomena \cite{Haberman:1979aa}. We will also relate it to the KdV equation \eqref{eqn:kdv} later on. It admits a monotonous solution with zero  acceleration ($d^2X/dt^2 = 0$) at $t \rightarrow -\infty$ given by 
\bea
X = \sqrt{-t} + \frac{1}{8t^2}+O(t^{-9/2})+ O(\eta^{2/5}).\label{AS}
\eea

The derivation of the equations \eqref{OTK} assumed the scaling \eqref{scaleOTK} and assumed $\eta \rightarrow 0$ with all independent quantities of order unity. However, in the high spin limit, another small parameter exists: $\lambda =\sqrt{1-\tilde a^2}\rightarrow 0$. It turns out that the relevant scalings and equations differ depending whether $\eta \gg \lambda $ or $\lambda \ll  \eta$ as we now detail. 

\section{The high spin limit of the Ori-Thorne-Kesden equations}

In the high spin limit, the Taylor coefficients are given up to $O(\lambda^{4/3})$ corrections by 
\bea
\alpha_* &=& 1 - 4 \cdot 2^{1/3} \lambda^{2/3},\quad c_* = 6\sqrt{3} \left( 1 - \frac{61}{12}\cdot 2^{1/3} \lambda^{2/3} \right),\nonumber \\
\beta_* &=& \frac{\sqrt{3} \lambda^{2/3}}{2^{2/3}},\quad d_* = -\frac{4}{\sqrt{3}} \left( 1 - \frac{17}{4} \cdot 2^{1/3} \lambda^{2/3} \right) ,\label{coefs}\\
\gamma_* &=& \sqrt{3} \left( 1- \frac{19}{2} \frac{\lambda^{2/3}}{2^{2/3}} \right),\quad \delta_* = \frac{4 \cdot 2^{1/3}}{\sqrt{3}}\lambda^{2/3}\nonumber .
\eea
Assuming the scaling \eqref{scaleOTK}, the equations of motion are given at leading order in $\lambda$ and $\eta$ by the high spin limit of the Ori-Thorne-Kesden equations
\bea
\frac{d^2 \tilde R}{d\tau^2} &=& - \tilde R^2  - \frac{\sqrt{3}}{2^{2/3}} \kappa_* \eta \lambda^{2/3} (\tau - \tau_*), \\
\frac{d(\chi - \xi)}{d\tau} &=& 2 \sqrt{3} \sigma_* \eta \tilde R.
\eea

We will now reformulate these equations. Remember that in the high spin regime, the Kerr metric can be written close to the ISCO as the near-horizon extreme Kerr (NHEK) metric with $O(\lambda^{1/3})$ corrections \cite{Bardeen:1999px}, see \cite{Bredberg:2011hp,Compere:2012jk,Compere:2017hsi} for reviews: 
\bea
&ds^2 &= M^2(1+\cos^2\theta) \Big( -R_N^2 dT_N^2 + \frac{dR_N^2}{R_N^2} + d\theta^2 \nonumber\\
&&\hspace{-15pt}+ \frac{4\sin^2\theta}{(1+\cos^2\theta)^2} (d\Phi_N + R_N dT_N)^2  \Big) +O(\lambda^{1/3}).\label{NHEK}
\eea 
The near-horizon coordinates $(T_N,R_N,\theta,\Phi_N)$ are related to the Boyer-Lindquist coordinates $(\tilde t,\tilde r,\theta,\tilde \phi)$ by 
\bea
\tilde t &=& 2M \lambda^{-2/3} T_N, \\ 
\tilde r &=& 1 + \lambda^{2/3} R_N, \\
\tilde \phi &=& \Phi_N +\lambda^{-2/3} T_N. 
\eea
The ISCO is located at $R_N=R_N^* \equiv 2^{1/3}+O(\lambda^{1/3})$. The NHEK energy $e_N$ and radius $R_N$ are therefore related to the asymptotically flat energy and Boyer-Lindquist radius as 
\bea
\tilde R &=& (R_N -R_N^*)\lambda^{2/3}, \nonumber \\
\tilde e &=& \frac{\ell}{2} + \frac{\lambda^{2/3}}{2}e_N.\label{OTs}
\eea
This implies 
\bea
\chi - \xi &=& (e_N + \frac{3}{4} 2^{1/3} \xi )\lambda^{2/3} .\label{def1}
\eea
In terms of the NHEK radius and energy, the high spin Ori-Thorne-Kesden equations can be therefore reformulated as 
\bea
\frac{d^2 R_N}{d\tau^2} &=&-\lambda^{2/3} (R_N-R_N^*)^2\nonumber \\
&& -4 \times 2^{1/3} \sigma_* \eta (\tau -\tau_*) , \label{eqHIGH1c}\\
\frac{de_N}{d\tau} &=& 2 \sqrt{3} \sigma_* \eta R_N.\label{eqHIGH2}
\eea

So far, we obtained these equations assuming the original scaling \eqref{scaleOTK}. More generally, in the presence of a small parameter $\lambda$ in addition to $\eta$, we need to specify how $\eta$ scales with $\lambda$. We define $\eps$ such that 
\bea
\eta \sim \lambda^{1+\eps}.\label{scaleeta}
\eea
There are three possible cases: 
\begin{itemize}
\item Standard high spin: $\eta \ll \lambda$ ($\eps > 0$);
\item Marginal high spin: $\eta \sim \lambda$ ($\eps = 0$);
\item Extremely high spin: $\lambda \ll \eta$  ($-1< \eps < 0$).
\end{itemize}
In standard astrophysical settings, the spin obeys the Thorne bound $\lambda > 0.06$ while $\eta$ can be extremely small for extreme mass ratio inspirals. It is therefore natural to consider the standard high spin scaling in order to model astrophysical scenarios. When $\eps \rightarrow \infty$, we will recover the Ori-Thorne-Kesden equations as we will detail below. For $\eps > 0$ finite, we can substitute the coefficients \eqref{coefs} and the redefinition \eqref{def1} into \eqref{master}. In order to take the limit $\lambda \rightarrow 0$ of \eqref{master} we need to specify the scaling of $R_N$, $e_N$ and $\tau$ in terms of $\eta$. The scaling of $\xi$ is then deduced from \eqref{linl}. We find that  the equations \eqref{eqHIGH1c}-\eqref{eqHIGH2} are invariant under the following scaling
\begin{subequations}\label{scaleHIGH2} \begin{align}
\tau -\tau_* & \sim \eta^{-\frac{1}{5}-\frac{2}{15(1+\eps)}},\\
 R_N - R_N^* & \sim  \eta^{\frac{2}{5}-\frac{2}{5(1+\eps)}}, \\
e_N \sim   \xi &\sim  \eta^{\frac{4}{5}- \frac{2}{15(1+\eps)}}, \\
e_N + \frac{3}{4} 2^{1/3} \xi & \sim  \eta^{\frac{6}{5}-\frac{8}{15(1+\eps)}}. 
\end{align}\end{subequations}
We now check that  \eqref{eqHIGH1c}-\eqref{eqHIGH2} can be also obtained from the full equations \eqref{accR}, \eqref{geo2R} using the scaling \eqref{scaleeta}-\eqref{scaleHIGH2}. This justifies the neglected terms in the Taylor expansion \eqref{master}. The Ori-Thorne scaling \eqref{scaleOTK} is recovered in the limit $\eps \rightarrow \infty$. We have therefore obtained the Ori-Thorne-Kesden equations in the presence of two small parameters $\eta,\lambda$ assuming $\eta \ll \lambda$.

\section{Transition equations in the marginal and extremely high spin limit}

We found that the scaling \eqref{scaleHIGH2} applies for all $\eps >0$. In the marginal high spin case $\eps = 0 $ of \eqref{scaleeta}, namely $\eta \sim \lambda$, we also find that the scaling solution \eqref{scaleHIGH2} holds. It reads explicity as
\bea
\tau -\tau_* & \sim &\eta^{-1/3},\qquad R_N - R_N^* \sim \eta^{0},\nonumber \\
e_N &\sim  &\xi \sim \eta^{2/3},  \label{scaleHIGH1}
\eea
or, equivalently, 
\bea
\tau -\tau_*  \sim \eta^{-1/3},\qquad \tilde R  \sim \eta^{2/3},\qquad \chi - \xi  \sim   \eta^{4/3}.  \label{scaleHIGH1bis}
\eea
 The energy equation \eqref{eqHIGH2} is unchanged while the radial equation (invariant under \eqref{scaleHIGH1}) gets modified to the ``marginal extremely high spin radial transition equation'' 
\bea
\frac{d^2 R_N}{d\tau^2} &=&-\lambda^{2/3}(R_N-R_N^*)^2 \nonumber\\
&&   -8 \sigma_* \eta (\tau - \tau_*) R_N+ \frac{2}{\sqrt{3}}e_N .\label{eqHIGH1b}
\eea
In the extremely high spin case $\lambda \ll \eta$ ($-1 < \eps <0$), we find that the scaling \eqref{scaleHIGH1bis} remains consistent. It can also be written as 
\bea
\tau -\tau_* & \sim &\eta^{-\frac{1}{3}},\qquad R_N - R_N^* \sim \eta^{\frac{2}{3}-\frac{2}{3(1+\eps)}},\nonumber \\
e_N &\sim  & \eta^{\frac{4}{3}-\frac{2}{3(1+\eps)}}, \qquad \xi \sim \eta^{\frac{2}{3}}. \label{scaleHIGH3}
\eea
The radial evolution is now governed by the  ``extremely high spin radial transition equation'' 
\bea
\frac{d^2 R_N}{d\tau^2} &=&-\lambda^{2/3}(R_N-R_N^*)^2 + 4 \eta \sigma_*  (\tau - \tau_*) R_N^* \nonumber\\
&&   -8 \sigma_* \eta (\tau - \tau_*) R_N+ \frac{2}{\sqrt{3}}e_N, \label{eqHIGH1a}
\eea
and the equation of motion for the energy is still 
\bea
\frac{de_N}{d\tau} &=& 2 \sqrt{3} \sigma_* \eta R_N.\label{eqHIGH2bis}
\eea
We checked that these equations can be also obtained from the full equations \eqref{accR}, \eqref{geo2R} using the scaling \eqref{scaleHIGH3}, which justifies the neglected terms in the Taylor expansion \eqref{master}. 

The extremely high spin radial transition equation \eqref{eqHIGH1a} together with the energy equation \eqref{eqHIGH2bis} can only be partly recovered from the NHEK geometry. If we assume the more restricted scaling regime (that probes a smaller range around the ISCO)
\bea
\tau -\tau_*  &\sim &\eta^{-1/3},\qquad R_N - R_N^* \sim \eta^{0},\nonumber \\
e_N &\sim  & \eta^{2/3}, \qquad \xi \sim \eta^{2/3}, \label{scaleHIGH4}
\eea
the energy equation remains unchanged while the radial equation \eqref{eqHIGH1a} simplifies to the ``NHEK transition equation''
\bea
\frac{d^2 R_N}{d\tau^2} &=& -8 \sigma_* \eta (\tau - \tau_*) R_N+ \frac{2}{\sqrt{3}}e_N .\label{eqHIGH1d}
\eea
This equation can be obtained from NHEK  as we now show. Since $\lambda \ll \eta$, we can take the formal limit $\lambda \rightarrow 0$ first and the $O(\lambda^{1/3})$ corrections to the NHEK geometry \eqref{NHEK} are negligible. In the NHEK geometry, the radial geodesic equations read as 
\bea
\left( \frac{dR_N}{d\tau} \right)^2 &=&  e_N^2 - V_N(R_N , e_N, \ell) ,\label{geoRN} \\
\frac{d^2R_N}{d\tau^2} &=& -\frac{1}{2}\frac{\p V_N(R_N ,e_N, \ell)}{\p R_N},
\eea
where $ e_N=2 \lambda^{-2/3} \tilde e - \lambda^{-2/3}\ell$ is the energy per unit probe mass per unit mass $M$ associated with $\p_{T_N}$. 
Here the radial potential takes a simple exact form,
\bea
V_N(R_N,e_N,\ell) &=& -2 e_N \ell R_N +(1-\frac{3}{4}\ell^2)R_N^2. 
\eea
These equations are compatible given the following constraint equation is obeyed
\bea
\frac{de_N}{d\tau} \frac{\p (V_N-e_N^2)}{\p e_N} + \frac{d \ell}{d\tau} \frac{\p V_N}{\p \ell} = 0.  
\eea
Up to $O(\lambda^{1/3})$ corrections, the evolution equations are therefore given by
\bea\label{eqfullNHEK}
\frac{d^2 R_N}{d\tau^2} &=& e_N \ell + (\frac{3}{4}\ell^2 - 1)R_N,\\
\frac{de_N}{d\tau} &=& \kappa_* \eta R_N \frac{e_N + \frac{3}{4} \ell R_N}{e_N + \ell R_N}.
\eea
The ISCO has energy,  angular momentum and radius $e_N =e_N^* \equiv 0 $, $\ell = \ell^* \equiv \frac{2}{\sqrt{3}}+O(\lambda^{2/3})$ and $R_N = R_N^* \equiv 2^{1/3}$. In the linear approximation in $e,\ell$ around the ISCO values and using \eqref{linl}, we finally obtain the NHEK transition equations \eqref{eqHIGH1d}-\eqref{eqHIGH2bis}.

It is important to note that we cannot obtain either \eqref{eqHIGH1c},  \eqref{eqHIGH1b} nor \eqref{eqHIGH1a} from motion in the NHEK geometry since the subleading corrections in $\lambda$ play a key role in the evolution. This points to a limitation of the near-horizon methods that use only the leading order near-horizon geometry \eqref{NHEK}. In this case, we can only recover a small subregion of the dynamics \eqref{eqHIGH1a} when $\lambda \ll \eta$.

Let us finally discuss the solutions to the equations \eqref{eqHIGH1b}, \eqref{eqHIGH1a}, \eqref{eqHIGH1d}. The NHEK transition equations can be easily solved analytically. Taking the derivative of \eqref{eqHIGH1d} with respect to proper time and substituting \eqref{eqHIGH2bis} we obtain the closed-form normalized third-order equation 
\bea
\frac{d^3 R_N}{dt^3} + (t - t_*) \frac{dR_N}{dt} + \frac{1}{2}R_N = 0,\label{trm}
\eea
after defining the rescaled time as $t = (8 \sigma_* \eta)^{1/3} \tau$, $t_* =  (8 \sigma_* \eta)^{1/3} \tau_*$. This equation admits analytic solutions in terms of Airy functions  
\bea
R_N &= & c_1 \text{Ai}(\frac{t_*-t}{2^{2/3}})^2 + c_2 \text{Ai}(\frac{t_*-t}{2^{2/3}}) \text{Bi}(\frac{t_*-t}{2^{2/3}}) \nn\\
&& +c_3 \text{Bi}(\frac{t_*-t}{2^{2/3}})^2,\label{solex}
\eea 
with three real integration constants $c_1,c_2,c_3$. Despite this simplicity and the fact that it can be recovered from NHEK, we will see later that \eqref{eqHIGH1d} cannot be connected to a quasi-circular inspiral and, more fundamentally,  the scaling regime where $\lambda \ll \eta$ is unphysical due to backreaction on the spin of the black hole.  

Less obvious is that \eqref{eqHIGH1b} relates to a well known, integrable nonlinear differential equation. With the same procedure as before, combining \eqref{eqHIGH1b}, \eqref{eqHIGH2bis} gives  
\bea
\frac{d^3 R_N}{dt^3} &+& \frac{1}{2}(\frac{\lambda}{\sigma_* \eta})^{2/3} (R_N-R_N^*) \frac{dR_N}{dt} \nonumber \\
&& + (t - t_*) \frac{dR_N}{dt} + \frac{1}{2}R_N = 0.
\label{eqn:kdvintermed1}
\eea
Similarly, for $\lambda \ll \eta$, combining \eqref{eqHIGH1a}, \eqref{eqHIGH2bis}, one finds instead
\bea
\frac{d^3 R_N}{dt^3} &+ & \frac{1}{2}(\frac{\lambda}{\sigma_* \eta})^{2/3} (R_N-R_N^*) \frac{dR_N}{dt} \nn \\
& & + (t -  t_*) \frac{dR_N}{dt} + \frac{1}{2}(R_N-R_N^*)= 0.
\label{eqn:kdvintermed2}
\eea
We will continue the analysis only for \eqref{eqn:kdvintermed1} but the result can immediately be applied to \eqref{eqn:kdvintermed2} by replacing $R_N$ with $R_N-R_N^*$ and $t_*$ with  $t_* - (\frac{\lambda}{2\sigma_* \eta})^{2/3}$. We now make the following substitutions
\bea
t - t_* &=& (\frac{2}{9}\zeta+2^{-2/3})(\frac{\lambda}{\sigma_* \eta})^{2/3}, \nn \\
R_N &=& \frac{2}{3}\psi(\zeta) -\frac{2}{3}\zeta,\nn \\
\omega^2 &=& \frac{243}{4} (\frac{\sigma_* \eta}{\lambda})^2,
\eea
to obtain
\be
-\frac{2}{3} \psi -\frac{1}{3} \zeta  \frac{d\psi}{d\zeta}+\psi \frac{d\psi}{d\zeta} + \omega^2 \frac{d^3\psi}{d\zeta^3} = 0. \label{eqn:selfsimkdv}
\ee
This equation can be obtained from the Korteweg-de Vries (KdV) equation for $\Psi(s,z)$ \cite{ablowitz1991solitons}
\be
\frac{\p \Psi}{\p s} + \Psi \frac{\p \Psi}{\p z} + \omega^2 \frac{\p^3 \Psi}{\p z^3} = 0,
\label{eqn:kdv}
\ee
by using the self similar variables
\be
\zeta = \frac{z}{s^{1/3}}, \qquad \psi(\zeta) = s^{2/3} \Psi(s,z).
\ee
Note that the first Painlev\'e equation can also be obtained as a particular reduction of the KdV equation \cite{ablowitz1991solitons}. Indeed set 
	\be
	\Psi(s,z) = (8 \omega^2)^{1/5}X(t)+ s, \; t = 2(8\omega^2)^{-2/5}(z-\frac{s^2}{2}).
	\ee
The equation following from \eqref{eqn:kdv} can then be integrated to find \eqref{OTK}.

\section{Inspiral in the near-horizon region}
 
The analytic quasi-circular adiabatic evolution in the near-horizon geometry was found in \cite{Gralla:2016qfw}. We are interested in the region between the ISCO and the boundary of the near-horizon region, $2^{1/3} < R_N \ll \lambda^{-2/3}$. The analysis starts with the result for the gravitational wave emission on a circular orbit in the near-horizon region of Kerr. The total energy radiated per unit Boyer-Lindquist time per probe mass is at leading order in $\lambda$ given by  \cite{Porfyriadis:2014fja,Gralla:2015rpa}
\bea
-\frac{de}{d (\tilde t/M)} = \sigma_* \eta\, (\tilde r-1). 
\eea
In terms of near-horizon variables, $\tilde r= 1+\lambda^{2/3}R_N$ and at leading order in $\lambda$, the latter equation is equivalent to 
\bea
\frac{d\ell}{d T_N} = -4 \sigma_* \eta R_N.\label{rell}
\eea
For a circular geodesic, the angular momentum and energy are given by
\bea
\ell_{circ} &=& \frac{2}{\sqrt{3}}+ \frac{4 (1+R_N^3)}{3 \sqrt{3} R_N^2}\lambda^{2/3}+O(\lambda^{4/3}),\label{lcN} \\
e_N^{circ} &=& -  \frac{4+R_N^3}{2 \sqrt{3} R_N}\lambda^{2/3}+O(\lambda^{4/3})\label{ecN}.
\eea
At leading order in $\lambda$, a circular geodesic in NHEK obeys 
\bea
\frac{dT_N}{d\tau} &=& \frac{e_N}{R_N^2}  +\frac{\ell}{R_N}  = \frac{2}{\sqrt{3} R_N} + O(\lambda^{2/3})\label{circT}.\label{circT0}
\eea
The two equations \eqref{rell}-\eqref{circT} lead to the linear decay of angular momentum per unit proper time \eqref{linl}, which is also assumed more generally in the transition region. Substituting \eqref{lcN} into \eqref{rell} we obtain
\bea
\frac{dR_N}{dT_N} =  -\frac{1}{T_N^\text{in}} \frac{R_N}{1-2 R_N^{-3}},\quad T^\text{in}_N \equiv \frac{\lambda^{2/3}}{3 \sqrt{3} \sigma_* \eta}.\label{dRdT}
\eea
The explicit solution is given by 
\bea
R_N(T_N) = R_0 e^{-\frac{T_N}{T^\text{in}_N}}e^{\frac{k}{3}+\frac{1}{3} W(-k e^{3\frac{T_N}{T^\text{in}_N} -k})},\;\; k\equiv \frac{2}{R_0^3},\label{Lambert}
\eea
where $W$ is the Lambert or product log function that obeys $W(0)=0$ and $W(-1/e)=-1$. At radii far from the ISCO but still in the near-horizon region, $2^{1/3} \ll R_N \ll \lambda^{-2/3}$, the $k$ terms are negligible and we  recognize the exponential decay of the early ($-T_N \gg T_N^\text{in}$) near-horizon quasi-circular adiabatic inspiral.

Combining \eqref{circT0} and \eqref{dRdT} we also obtain 
\bea
(1-\frac{2}{R_N^3})\frac{dR_N}{d\tau} = -\frac{1}{\tau^{\text{in}}},\qquad \tau^\text{in} \equiv \frac{\sqrt{3}}{2} T_N^\text{in} =\frac{\lambda^{2/3}}{6 \sigma_* \eta} 
\eea
which we can integrate to obtain
\bea
R_N+\frac{1}{R_N^2}-R_0-\frac{1}{R_0^2} = \frac{\tau_0 - \tau}{\tau^\text{in}}\label{inspsol}
\eea
where $R_N(\tau_0) = R_0$. The explicit solution for $R_N(\tau)$ is the only real cubic root of this equation. Evaluating at $\tau_*$, the proper time elapsed between the initial time and reaching the ISCO is \footnote{This equation can also be found directly by evaluating \eqref{linl} at $\tau = \tau_0$ and using the fact that the initial angular momentum is given by the circular geodesic expression \eqref{lcN}.}
\bea
\tau_* - \tau_0 = \tau^{\text{in}} \left( R_0 + R_0^{-2} -\frac{3}{2^{2/3}} \right). \label{t0ts}
\eea
It is straightforward to obtain that the inspiral \eqref{inspsol} admits around the ISCO the behavior
\bea
R_N -R_N^* = \frac{2^{2/3}}{\sqrt{3}} \frac{\sqrt{\tau_{*}-\tau}}{\sqrt{\tau^{\text{in}}}}+O(\tau_*-\tau). \label{ci}
\eea

\section{Spin evolution during the inspiral}

Before presenting the matching with the transition, it is important to first establish to what extent the probe will spin down the central black hole due to superradiance and, more precisely, whether a binary system with $\lambda \ll \eta$ can exist. Heuristic formulae for the spin evolution were derived in \cite{1970Natur.226...64B,Hughes:2002ei,Buonanno:2007sv,Kesden:2008ga,Kesden:2009ds}. Based on these heuristics, it was shown that superradiant scattering slows down the central black hole during the inspiral \cite{Kesden:2009ds}. A fundamental derivation that takes into account the specificities of the near-extremal behavior is however necessary to quantify this slow down. Such fundamental analysis was performed in \cite{Gralla:2016qfw} using the explicit analytical quasi-circular inspiral evolution but an error occured on Eq. (23). We will therefore restart the analysis here. 

So far we assumed that $M$ and $J$ (and therefore $\lambda = \sqrt{1-J^2/M^4}$) are constant, but in fact they evolve already at first order in the self-force (which is proportional to $\mu$). We would like to derive a lower bound on $\lambda$. Let us assume that the black hole is exactly extremal ($\lambda =0$) when the compact object enters the near-horizon region. We further assume that it follows the quasi-circular inspiral \eqref{Lambert} or, equivalently, \eqref{inspsol}, at the entry of the near-horizon region. Thanks to the gravitational wave absorption in the near-horizon region, the central black hole mass evolves as
\bea
\frac{dM}{d\tilde t} = \sigma_{*,H} \eta^2 (\tilde r - 1).
\eea
In this case, $\sigma_{*,H} < 0$, see \eqref{sigma}, and energy is extracted from the black hole. This is a consequence of superradiance. The angular momentum obeys
\be
\frac{dJ}{d\tilde t} = \tilde \Omega^{-1}\frac{dM}{d\tilde t},\quad \tilde\Omega =\frac{1}{2M}\left(1-\frac{3}{4}\lambda^{2/3}R_N\right)+O(\lambda). 
\ee
This leads to the departure from extremality
\bea
\frac{d\lambda}{d\tilde t} = -\frac{3\lambda^{1/3}}{2M}  \sigma_{*,H} \eta^2 R_N^2 >0. 
\eea
Using $\tilde t = 2 M \lambda^{-2/3}T_N$ and \eqref{circT0}, the evolution of $\lambda$ is given in terms of proper time as
\bea
\lambda \frac{d\lambda}{d\tau} =-2\sqrt{3}\sigma_{*,H} \eta^2 \,\lambda^{2/3} R_N. 
\eea
At the entry of the near-horizon region $R_N = R_0 \sim \lambda^{-2/3}$. This implies that $\lambda^2$ will develop a linear growth proportional to $\eta^2$ or $\lambda \sim \eta$, at least. The start of the inspiral evolution therefore rules out extremely high spin binaries with $\lambda \ll \eta$. The marginal scaling $\lambda \sim \eta$ is not ruled out by this argument.  

\section{Match between the inspiral and the transition}

In the standard high spin case $\eta \ll \lambda$, the transition motion is given by the high spin limit of the Ori-Thorne-Kesden equations \eqref{eqHIGH1c}-\eqref{eqHIGH2} while the physical quantities scale as \eqref{scaleHIGH2}. The behavior of the inspiral close to the ISCO is given by \eqref{ci}. It exactly corresponds to the asymptotic quasi-circular inspiral condition that cancels at leading order the right-hand side of \eqref{eqHIGH1c}. We can therefore match the transition solution \eqref{AS} to the $\tau \rightarrow -\infty$ asymptotics \eqref{ci}. The matching can be performed uniquely for any given set of physical parameters $\eta$, $\lambda$ with $\eta \ll \lambda$.

The result is depicted on Figure 1 for $\eta = 10^{-6}$, $\lambda = 10^{-3}$. We fixed the ambiguity in shifting the proper time by choosing $\tau_* = 0$ (i.e. $\tau = 0$ corresponds to $\ell = \ell_*$). The dotted blue curve is the matching solution
\bea
R_N =R_N^*+ \frac{2^{2/3}}{\sqrt{3}} \frac{\sqrt{\tau_{*}-\tau}}{\sqrt{\tau^{\text{in}}}}\label{matchR}
\eea
that asymptotes at $\tau \rightarrow 0$ to the inspiral \eqref{ci} and at $\tau \rightarrow -\infty$ to the transition solution \eqref{AS}. 

The plunge always occurs at $\tau > \tau_*$. The final plunge is therefore a subcritical geodesic in the sense that $\ell < \ell_*$. In \cite{Compere:2017hsi}, only critical ($\ell = \ell_*$) and supercritical $(\ell > \ell_*)$ geodesics were described since only geodesics that enter the NHEK region from radial infinity were considered. Instead, subcritical geodesics only exist up to a finite radius within the NHEK region. They can be obtained as analytic continuation of the supercritical orbits, as will be detailed in \cite{Compere:2019bb}, see also \cite{Kapec:2019hro}. 

For the marginal $\lambda \sim \eta$ and extremely high spin case $\lambda \ll \eta$ we find that the solution \eqref{matchR} is not an asymptotic solution at $\tau \rightarrow -\infty$ of the transition equations \eqref{eqHIGH1b} or \eqref{eqHIGH1a} combined with \eqref{eqHIGH2bis}. If one assumes such an ansatz, an overleading $\tau^{3/2}$ term appears which invalidates the ansatz. This compounds our earlier observation that the extremely high spin case $\lambda \ll \eta$ is inconsistent with the spin evolution. 

Even in the restricted scaling \eqref{scaleHIGH4} where the NHEK metric is sufficient to describe the motion, the NHEK transition equations are inconsistent with the quasi-circular near-horizon inspiral. Indeed, the NHEK transition equations \eqref{eqHIGH1d}-\eqref{eqHIGH2bis} do not admit an asymptotic solution without acceleration, which would have been an analogue of \eqref{AS}. Explicitly, imposing that the right-hand side of \eqref{eqHIGH1d} be zero at $\tau \rightarrow -\infty$ leads to $R \sim (-\tau)^{-1/2}$ which is inconsistent since $R$ has to be both positive and monotonous. We deduce that no solution \eqref{solex} can be matched into the adiabatically evolving quasi-circular inspiral near the ISCO \eqref{ci} at $\tau \rightarrow -\infty$. \footnote{This can also be explained by the following argument. The NHEK transition dynamics  is defined in terms of finite $\lambda^0$ quantities within the near-horizon geometry. In contrast, the quasi-circular near-horizon inspiral is driven by $\lambda^{2/3}$ corrections (since all circular orbits have the same energy in the near-horizon region at order $\lambda^0$). There is therefore no possible match between these two motions.}

\section{Conclusion}

We obtained three sets of non-quasi-circular transition equations in the near-horizon region of the near-extremal Kerr black hole at leading order in the high spin limit that depend upon the relative scaling between the near-extremality parameter $\lambda$ and the mass ratio $\eta$. The Ori-Thorne-Kesden transition equations \cite{Ori:2000zn,Kesden:2011ma} were derived for arbitrary parameters $\eta \ll \lambda \ll 1$ and the radial evolution equation was identified as the Painlev\'e transcendent equation of the first kind. For these Ori-Thorne-Kesden transition equations, we found a unique match with the near-horizon inspiral of \cite{Gralla:2016qfw}. In contrast, we obtained transition equations for marginal $\eta \sim \lambda \ll 1$ and extremely high spin scalings $\lambda \ll \eta \ll 1$ that can be written as the KdV equation with self-similar variables. However,  we showed that they cannot be matched to the near-horizon quasi-circular inspiral of \cite{Gralla:2016qfw}. 

In fact, we showed that the spin evolution during the near-horizon inspiral drives the spin of the central black hole away from extremality as $\lambda \sim \eta$ or higher due to superradiant gravitational wave extraction of angular momentum. This rules out the scaling $\lambda \ll \eta$ which compounds the absence of a match between the transition equations and the quasi-circular inspiral for that range of parameters. 

While the near-horizon limit was instrumental in formulating the inspiral and transition motion in the high spin limit, we found that the NHEK metric alone is insufficient to describe the transition motion even in the high spin regime. Corrections to the near-horizon metric originating from both the deviation from maximal spin and from the backreaction of the probe play an essential role in the transition motion of extreme mass ratio coalescences. 


\section*{Acknowledgments}
We thank Samuel Gralla for the suggestion to look at this problem. We thank Matthias Fabry, Thomas Hertog, and Andrew Strominger for useful discussions. We thank Ollie Burke, Jonathan Gair and Joan Sim\'on for sharing their draft. G.C. is research associate of the F.R.S.-FNRS. G.C. also acknowledges support from the IISN convention 4.4503.15 and networking support from the COST Action GWverse CA16104. K.F. is aspirant van het Fonds Wetenschappelijk Onderzoek - Vlaanderen.

\appendix
\vspace{-10pt}

\section{Circular geodesics of Kerr}
\label{app:circ}

The angular velocity of circular orbits in Boyer-Lindquist coordinates is given by
\bea
\tilde \Omega = (\tilde r^{3/2} + \tilde a)^{-1}.
\eea
The angular momentum per unit probe mass and central black hole mass and energy per unit probe mass is given for circular orbits by 
\bea
\ell_{circ}(\tilde r ; \lambda)&=& \frac{\sqrt{\tilde r}-2 \tilde a / \tilde r + \tilde a^2 / \tilde r^{3/2}}{\sqrt{1-3/\tilde r + 2\tilde a / \tilde r^{3/2}}} , \\
\tilde e_{circ}(\tilde r ; \lambda) &=& \frac{1-2/\tilde r +\tilde a / \tilde r^{3/2}}{\sqrt{1-3/\tilde r + 2\tilde a / \tilde r^{3/2}} },
\eea
where $\tilde a = \sqrt{1-\lambda^2}$. The ISCO is located at 
\bea
\tilde r_* &=& 3+Z_2 - \text{sign}(\tilde a) [(3-Z_1)(3+Z_1+2Z_2) ]^{1/2}, \nonumber \\ 
&=& 1 + 2^{1/3}\lambda^{2/3}+O(\lambda^{4/3})
\eea
where 
\bea
Z_1 &\equiv & 1+(1-\tilde{a}^2)^{1/3}[ (1+\tilde a)^{1/3}+(1-\tilde a)^{1/3}], \\
 Z_2 &\equiv &(3\tilde a^2 + Z_1^2)^{1/2}.\nonumber
\eea
At the ISCO, the angular velocity is 
\bea
\tilde \Omega_* = \frac{1}{\tilde r_*^{3/2}+\tilde a}= \frac{1}{2}-\frac{3 \lambda^{2/3}}{4 \times 2^{2/3}}+O(\lambda^{4/3})
\eea
and the probe angular momentum and Boyer-Lindquist energy are 
\bea
\ell_{*} &=& \frac{6\sqrt{\tilde r_*} -4 \tilde a}{\sqrt{3\tilde r_*}}  = \frac{2}{\sqrt{3}}+ \frac{2^{4/3}}{\sqrt{3}}\lambda^{2/3}+O(\lambda^{4/3}),\label{lcNstar} \\
\tilde e_* &=& \frac{1-2/\tilde r_* +\tilde a / \tilde r_*^{3/2}}{\sqrt{1-3/\tilde r_* + 2\tilde a / \tilde r_*^{3/2}} } ,\\
&=&  \frac{1}{\sqrt{3}}+ \frac{2^{1/3}}{\sqrt{3}}\lambda^{2/3}+O(\lambda^{4/3})
\eea
\vspace{10pt}

\providecommand{\href}[2]{#2}\begingroup\raggedright\endgroup


\begin{thebibliography}{10}
	
	\bibitem{Audley:2017drz}
	{\bf LISA} Collaboration, P.~Amaro-Seoane {\em et al.}, ``{Laser Interferometer
		Space Antenna},''
	\href{http://www.arXiv.org/abs/1702.00786}{{\tt 1702.00786}}.
	
	\bibitem{AmaroSeoane:2007aw}
	P.~Amaro-Seoane, J.~R. Gair, M.~Freitag, M.~Coleman~Miller, I.~Mandel, C.~J.
	Cutler, and S.~Babak, ``{Astrophysics, detection and science applications of
		intermediate- and extreme mass-ratio inspirals},'' {\em Class. Quant. Grav.}
	{\bf 24} (2007) R113--R169,
	\href{http://www.arXiv.org/abs/astro-ph/0703495}{{\tt astro-ph/0703495}}.
	
	\bibitem{Baibhav:2019rsa}
	V.~Baibhav {\em et al.}, ``{Probing the Nature of Black Holes: Deep in the mHz
		Gravitational-Wave Sky},''
	\href{http://www.arXiv.org/abs/1908.11390}{{\tt 1908.11390}}.
	
	\bibitem{Miller:2004va}
	M.~C. Miller, ``{Probing general relativity with mergers of supermassive and
		intermediate-mass black holes},'' {\em Astrophys. J.} {\bf 618} (2004)
	426--431,
	\href{http://www.arXiv.org/abs/astro-ph/0409331}{{\tt astro-ph/0409331}}.
	
	\bibitem{Smith:2013mfa}
	R.~J.~E. Smith, I.~Mandel, and A.~Vechhio, ``{Studies of waveform requirements
		for intermediate mass-ratio coalescence searches with advanced
		gravitational-wave detectors},'' {\em Phys. Rev.} {\bf D88} (2013), no.~4,
	044010,
	\href{http://www.arXiv.org/abs/1302.6049}{{\tt 1302.6049}}.
	
	\bibitem{Ori:2000zn}
	A.~Ori and K.~S. Thorne, ``{The Transition from inspiral to plunge for a
		compact body in a circular equatorial orbit around a massive, spinning black
		hole},'' {\em Phys. Rev.} {\bf D62} (2000) 124022,
	\href{http://www.arXiv.org/abs/gr-qc/0003032}{{\tt gr-qc/0003032}}.
	
	\bibitem{Buonanno:2000ef}
	A.~Buonanno and T.~Damour, ``{Transition from inspiral to plunge in binary
		black hole coalescences},'' {\em Phys. Rev.} {\bf D62} (2000) 064015,
	\href{http://www.arXiv.org/abs/gr-qc/0001013}{{\tt gr-qc/0001013}}.
	
	\bibitem{Buonanno:2005xu}
	A.~Buonanno, Y.~Chen, and T.~Damour, ``{Transition from inspiral to plunge in
		precessing binaries of spinning black holes},'' {\em Phys. Rev.} {\bf D74}
	(2006) 104005,
	\href{http://www.arXiv.org/abs/gr-qc/0508067}{{\tt gr-qc/0508067}}.
	
	\bibitem{Damour:2007xr}
	T.~Damour and A.~Nagar, ``{Faithful effective-one-body waveforms of
		small-mass-ratio coalescing black-hole binaries},'' {\em Phys. Rev.} {\bf
		D76} (2007) 064028,
	\href{http://www.arXiv.org/abs/0705.2519}{{\tt 0705.2519}}.
	
	\bibitem{Sundararajan:2008bw}
	P.~A. Sundararajan, ``{The Transition from adiabatic inspiral to geodesic
		plunge for a compact object around a massive Kerr black hole: Generic
		orbits},'' {\em Phys. Rev.} {\bf D77} (2008) 124050,
	\href{http://www.arXiv.org/abs/0803.4482}{{\tt 0803.4482}}.
	
	\bibitem{Damour:2009kr}
	T.~Damour and A.~Nagar, ``{An Improved analytical description of inspiralling
		and coalescing black-hole binaries},'' {\em Phys. Rev.} {\bf D79} (2009)
	081503,
	\href{http://www.arXiv.org/abs/0902.0136}{{\tt 0902.0136}}.
	
	\bibitem{Kesden:2011ma}
	M.~Kesden, ``{Transition from adiabatic inspiral to plunge into a spinning
		black hole},'' {\em Phys. Rev.} {\bf D83} (2011) 104011,
	\href{http://www.arXiv.org/abs/1101.3749}{{\tt 1101.3749}}.
	
	\bibitem{Pan:2013rra}
	Y.~Pan, A.~Buonanno, A.~Taracchini, L.~E. Kidder, A.~H. Mrou\'e, H.~P.
	Pfeiffer, M.~A. Scheel, and B.~Szil\'agyi, ``{Inspiral-merger-ringdown
		waveforms of spinning, precessing black-hole binaries in the
		effective-one-body formalism},'' {\em Phys. Rev.} {\bf D89} (2014), no.~8,
	084006,
	\href{http://www.arXiv.org/abs/1307.6232}{{\tt 1307.6232}}.
	
	\bibitem{Taracchini:2014zpa}
	A.~Taracchini, A.~Buonanno, G.~Khanna, and S.~A. Hughes, ``{Small mass plunging
		into a Kerr black hole: Anatomy of the inspiral-merger-ringdown waveforms},''
	{\em Phys. Rev.} {\bf D90} (2014), no.~8, 084025,
	\href{http://www.arXiv.org/abs/1404.1819}{{\tt 1404.1819}}.
	
	\bibitem{Apte:2019txp}
	A.~Apte and S.~A. Hughes, ``{Exciting black hole modes via misaligned
		coalescences: I. Inspiral, transition, and plunge trajectories using a
		generalized Ori-Thorne procedure},'' {\em Phys. Rev.} {\bf D100} (2019),
	no.~8, 084031,
	\href{http://www.arXiv.org/abs/1901.05901}{{\tt 1901.05901}}.
	
	\bibitem{1974ApJ...191..507T}
	K.~S. {Thorne}, ``{Disk-Accretion onto a Black Hole. II. Evolution of the
		Hole},'' {\em Astrophys.J.} {\bf 191} (July, 1974) 507--520.
	
	\bibitem{1980AcA....30...35A}
	M.~A. {Abramowicz} and J.~P. {Lasota}, ``{Spin-up of black holes by thick
		accretion disks},'' {\em Acta Astron.} {\bf 30} (1980) 35--39.
	
	\bibitem{Bardeen:1999px}
	J.~M. Bardeen and G.~T. Horowitz, ``{The Extreme Kerr throat geometry: A Vacuum
		analog of AdS(2) x S**2},'' {\em Phys. Rev.} {\bf D60} (1999) 104030,
	\href{http://www.arXiv.org/abs/hep-th/9905099}{{\tt hep-th/9905099}}.
	
	\bibitem{Porfyriadis:2014fja}
	A.~P. Porfyriadis and A.~Strominger, ``{Gravity waves from the Kerr/CFT
		correspondence},'' {\em Phys. Rev.} {\bf D90} (2014), no.~4, 044038,
	\href{http://www.arXiv.org/abs/1401.3746}{{\tt 1401.3746}}.
	
	\bibitem{Hadar:2014dpa}
	S.~Hadar, A.~P. Porfyriadis, and A.~Strominger, ``{Gravity Waves from
		Extreme-Mass-Ratio Plunges into Kerr Black Holes},'' {\em Phys. Rev.} {\bf
		D90} (2014), no.~6, 064045,
	\href{http://www.arXiv.org/abs/1403.2797}{{\tt 1403.2797}}.
	
	\bibitem{Hadar:2015xpa}
	S.~Hadar, A.~P. Porfyriadis, and A.~Strominger, ``{Fast plunges into Kerr black
		holes},'' {\em JHEP} {\bf 07} (2015) 078,
	\href{http://www.arXiv.org/abs/1504.07650}{{\tt 1504.07650}}.
	
	\bibitem{Gralla:2015rpa}
	S.~E. Gralla, A.~P. Porfyriadis, and N.~Warburton, ``{Particle on the Innermost
		Stable Circular Orbit of a Rapidly Spinning Black Hole},'' {\em Phys. Rev.}
	{\bf D92} (2015), no.~6, 064029,
	\href{http://www.arXiv.org/abs/1506.08496}{{\tt 1506.08496}}.
	
	\bibitem{Gralla:2016qfw}
	S.~E. Gralla, S.~A. Hughes, and N.~Warburton, ``{Inspiral into Gargantua},''
	{\em Class. Quant. Grav.} {\bf 33} (2016), no.~15, 155002,
	\href{http://www.arXiv.org/abs/1603.01221}{{\tt 1603.01221}}.
	
	\bibitem{Hadar:2016vmk}
	S.~Hadar and A.~P. Porfyriadis, ``{Whirling orbits around twirling black holes
		from conformal symmetry},'' {\em JHEP} {\bf 03} (2017) 014,
	\href{http://www.arXiv.org/abs/1611.09834}{{\tt 1611.09834}}.
	
	\bibitem{Compere:2017hsi}
	G.~Comp{\`e}re, K.~Fransen, T.~Hertog, and J.~Long, ``{Gravitational waves from
		plunges into Gargantua},'' {\em Class. Quant. Grav.} {\bf 35} (2018), no.~10,
	104002,
	\href{http://www.arXiv.org/abs/1712.07130}{{\tt 1712.07130}}.
	
	\bibitem{Chen:2019hac}
	B.~Chen, G.~Comp{\`e}re, Y.~Liu, J.~Long, and X.~Zhang, ``{Spin and Quadrupole
		Couplings for High Spin Equatorial Intermediate Mass-ratio Coalescences},''
	{\em Class. Quant. Grav.} {\bf 36} (2019), no.~24, 245011,
	\href{http://www.arXiv.org/abs/1901.05370}{{\tt 1901.05370}}.
	
	\bibitem{1996PhRvD..53.3064R}
	F.~D. {Ryan}, ``{Effect of gravitational radiation reaction on nonequatorial
		orbits around a Kerr black hole},'' {\em Phys. Rev. D.} {\bf 53} (Mar., 1996)
	3064--3069, \href{http://www.arXiv.org/abs/gr-qc/9511062}{{\tt
			gr-qc/9511062}}.
	
	\bibitem{Kennefick:1998ab}
	D.~Kennefick, ``{Stability under radiation reaction of circular equatorial
		orbits around Kerr black holes},'' {\em Phys. Rev.} {\bf D58} (1998) 064012,
	\href{http://www.arXiv.org/abs/gr-qc/9805102}{{\tt gr-qc/9805102}}.
	
	\bibitem{Glampedakis:2002ya}
	K.~Glampedakis and D.~Kennefick, ``{Zoom and whirl: Eccentric equatorial orbits
		around spinning black holes and their evolution under gravitational radiation
		reaction},'' {\em Phys. Rev.} {\bf D66} (2002) 044002,
	\href{http://www.arXiv.org/abs/gr-qc/0203086}{{\tt gr-qc/0203086}}.
	
	\bibitem{Kapec:2019hro}
	D.~Kapec and A.~Lupsasca, ``{Particle motion near high-spin black holes},''
	{\em Class. Quant. Grav.} {\bf 37} (2020), no.~1, 015006,
	\href{http://www.arXiv.org/abs/1905.11406}{{\tt 1905.11406}}.
	
	\bibitem{Compere:2019bb}
	G.~Comp{\`e}re and A.~Druart, ``{Near-horizon Geodesics of High Spin Black
		Holes},''
	\href{http://www.arXiv.org/abs/2001.03478}{{\tt 2001.03478}}.
	
	\bibitem{Simon:2019aa}
	O.~Burke, J.~R. Gair, and J.~Sim{\'o}n, ``Transition from inspiral to plunge: A
	complete near-extremal waveform,'' {\em arXiv preprint arXiv:1909.12846}
	(2019).
	
	\bibitem{Pound:2005fs}
	A.~Pound, E.~Poisson, and B.~G. Nickel, ``{Limitations of the adiabatic
		approximation to the gravitational self-force},'' {\em Phys. Rev.} {\bf D72}
	(2005) 124001,
	\href{http://www.arXiv.org/abs/gr-qc/0509122}{{\tt gr-qc/0509122}}.
	
	\bibitem{Haberman:1979aa}
	R.~Haberman, ``{Slowly Varying Jump and Transition Phenomena Associated with
		Algebraic Bifurcation Problems},'' {\em SIAM J. Appl. Math.} {\bf 37(1)}
	(1979) 69--106.
	
	\bibitem{Bredberg:2011hp}
	I.~Bredberg, C.~Keeler, V.~Lysov, and A.~Strominger, ``{Cargese Lectures on the
		Kerr/CFT Correspondence},'' {\em Nucl. Phys. Proc. Suppl.} {\bf 216} (2011)
	194--210,
	\href{http://www.arXiv.org/abs/1103.2355}{{\tt 1103.2355}}.
	
	\bibitem{Compere:2012jk}
	G.~Comp{\`e}re, ``{The Kerr/CFT correspondence and its extensions},'' {\em
		Living Rev. Rel.} {\bf 15} (2012) 11,
	\href{http://www.arXiv.org/abs/1203.3561}{{\tt 1203.3561}}.
	[Living Rev. Rel.20,no.1,1(2017)].
	
	\bibitem{ablowitz1991solitons}
	M.~J. Ablowitz, M.~Ablowitz, P.~Clarkson, and P.~A. Clarkson, {\em Solitons,
		nonlinear evolution equations and inverse scattering}, vol.~149.
	\newblock Cambridge university press, 1991.
	
	\bibitem{1970Natur.226...64B}
	J.~M. {Bardeen}, ``{Kerr Metric Black Holes},'' {\em Nature} {\bf 226} (Apr.,
	1970) 64--65.
	
	\bibitem{Hughes:2002ei}
	S.~A. Hughes and R.~D. Blandford, ``{Black hole mass and spin coevolution by
		mergers},'' {\em Astrophys. J.} {\bf 585} (2003) L101--L104,
	\href{http://www.arXiv.org/abs/astro-ph/0208484}{{\tt astro-ph/0208484}}.
	
	\bibitem{Buonanno:2007sv}
	A.~Buonanno, L.~E. Kidder, and L.~Lehner, ``{Estimating the final spin of a
		binary black hole coalescence},'' {\em Phys. Rev.} {\bf D77} (2008) 026004,
	\href{http://www.arXiv.org/abs/0709.3839}{{\tt 0709.3839}}.
	
	\bibitem{Kesden:2008ga}
	M.~Kesden, ``{Can binary mergers produce maximally spinning black holes?},''
	{\em Phys. Rev.} {\bf D78} (2008) 084030,
	\href{http://www.arXiv.org/abs/0807.3043}{{\tt 0807.3043}}.
	
	\bibitem{Kesden:2009ds}
	M.~Kesden, G.~Lockhart, and E.~S. Phinney, ``{Maximum black-hole spin from
		quasi-circular binary mergers},'' {\em Phys. Rev.} {\bf D82} (2010) 124045,
	\href{http://www.arXiv.org/abs/1005.0627}{{\tt 1005.0627}}.
	
\end{thebibliography}
\end{document}